%% file: susy03_x.tex
\documentclass[aps,prl,preprint,groupedaddress,amssymb]{revtex4}
\usepackage{color}
\usepackage{epsfig}
\def\plb#1{Phys.~Lett.~{\bf B#1}}
\def\npb#1{Nucl.~Phys.~{\bf B#1}} 

\def\prd#1{Phys.~Rev.~{\bf D#1}}

\def\jhep#1{J. High Energy Phys.~{\bf #1}}

\begin{document}
\title{Unification of Couplings and Proton Decay \\
in SUSY GUTs
\footnote{
\uppercase{T}alk presented
at {\it \uppercase{SUSY} 2003:
\uppercase{S}upersymmetry in the \uppercase{D}esert}\/, 
held at the \uppercase{U}niversity of \uppercase{A}rizona,
\uppercase{T}ucson, \uppercase{AZ}, \uppercase{J}une 5-10, 2003.
\uppercase{T}o appear in the \uppercase{P}roceedings.}
}
%%  PLEASE USE THE ABOVE FOOTNOTE FOR ALL ARXIV POSTINGS,
%%   (SUBSTITUTING SPEAKER NAME, OF COURSE;  NO SPEAKER NAME NEEDED
%%      FOR SINGLE-AUTHOR CONTRIBUTIONS). 
%%  NOT NECESSARY TO HAVE FOOTNOTE FOR VERSIONS THAT ARE SENT TO 
%%  SUSY 2003 CONFERENCE FOR PUBLICATION

\author{ Radovan Derm\' \i\v sek}
\email[]{dermisek@physics.ucdavis.edu}

\affiliation{Davis Institute for High Energy Physics,
University of California, Davis, CA 95616, U.S.A.}

\date{January 14, 2004}

\begin{abstract}

We show that gauge couplings unify more precisely in the region of SUSY parameter space already 
preferred by Yukawa unification.
While proton decay due to dimension 5 operators is maximally suppressed in this region, 
the contribution from dimension 6 operators is enhanced as a 
consequence of lower unification scale.

\end{abstract}

\maketitle

\subsection{Introduction}

It has been pointed out that assuming high degree of universality in soft SUSY breaking terms
and
positive $\mu$ (preferred by $b \to s \gamma$ and the muon anomalous magnetic moment) 
top-bottom-tau Yukawa coupling unification 
(motivated by SO(10) symmetry) can be satisfied only in a narrow
region of SUSY parameter space~\cite{bdr}.
In that analysis the minimal SO(10) boundary conditions for soft SUSY breaking parameters at the 
GUT scale were considered:
universal squark and slepton masses -- $m_{16}$; universal gaugino masses -- $M_{1/2}$;
universal trilinear couplings -- $A$; and non-universal Higgs $H_u, \, H_d$ masses 
(non-universality in Higgs masses is well motivated by GUT scale threshold 
corrections~\cite{bdr}). The last two parameters can be exchanged for $\mu$ (SUSY Higgs mass) 
and $m_A$ (CP odd Higgs boson mass) by requiring radiative 
electroweak symmetry breaking with resulting values of $\mu$ and $m_A$. The region preferred by 
Yukawa coupling unification is specified by $\mu, \, M_{1/2} \ll m_{16}$, $A \sim -2 m_{16}$ and
$\tan \beta \sim 50 \pm 2$. The fit is improving and the region of allowed $\mu$ and $M_{1/2}$ 
grows with increasing $m_{16}$. These results can be understood by studying SUSY threshold 
corrections~\cite{bdr} and have been verified in independent analysis~\cite{yuk_other}.
The squark and slepton masses have to be quite heavy, $m_{16} \gtrsim 1.4$ TeV, preferably few 
TeV. Nevertheless, it was shown that these boundary conditions lead to maximal hierarchy between 
first two generations and the third generation due to RG running~\cite{Bagger:1999sy}. 
Therefore, taking $m_{16} \simeq 10$ TeV may still be considered natural, since the mass of the 
stop will be in a TeV region, roughly $m_{16}/10$. Masses of the first two generations of squarks 
and sleptons are close to $m_{16}$.

It is well known that heavier superpartners lead to more precise unification  
of gauge couplings~\cite{gcu_cor_cpp,gcu_cor_bmp,Feng:2000bp}.  
Since heavy superpartners are already required by Yukawa unification, it is 
interesting to see how gauge coupling unification improves in the same region 
of SUSY parameter space.

\subsection{Gauge coupling unification}

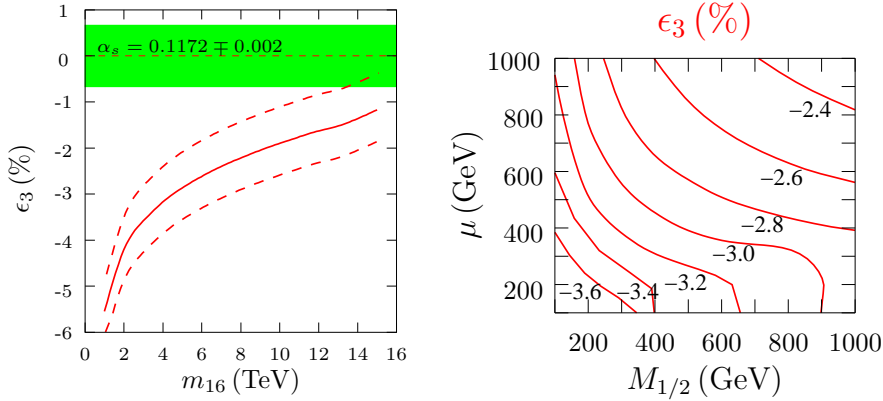
\begin{figure}[t]
\begin{center}
\input{eps_3_p_70.pstex_t}
\caption{
Value of $\epsilon_3$ in \% as a function of $m_{16}$ with all other
parameters
fixed as explained in the text (left); and as a function of ($\mu$, $M_{1/2}$)
for $m_{16} = 5$ TeV and $A_0 = -1.8 \; m_{16}$ (right).
}

\label{figure:1}
\end{center}
\end{figure}

Precise analysis of gauge coupling unification, including two-loop renormalization 
group equations, leading-log~\cite{gut_2loop} and finite~\cite{gcu_cor_cpp,gcu_cor_bmp} weak scale 
threshold
corrections,
together with improved  measurements of electroweak data reveals that 
gauge couplings miss each other with  a
significant discrepancy.
If we define the GUT scale as a scale at which the first two gauge 
couplings meet, $\alpha_1(M_G) = \alpha_2(M_G) \equiv  \alpha_G$,
%\begin{equation}
%\alpha_1(M_G) = \alpha_2(M_G) \equiv  \alpha_G
%\end{equation}
the value of the third gauge coupling at the weak scale can be predicted. 
With mSUGRA boundary conditions and requiring $m_0 < 2$ TeV the predicted $\alpha_s$ 
is $0.125 < \alpha_s (M_Z) < 0.143$~\cite{gcu_cor_cpp,gcu_cor_bmp,Feng:2000bp},
%:\cite{gcu_cor_cpp,gcu_cor_bmp,Feng:2000bp}
%\begin{equation}
% 0.125 < \alpha_s (M_Z) < 0.143
%\end{equation}
which should be compared with the current experimental value of the strong 
coupling constant, $\alpha_s (M_Z) = 0.1172 \pm 0.002$~\cite{pdg2002}. 
%:\cite{pdg2002}
%\begin{equation}
%\alpha_s (M_Z) = 0.1172 \pm 0.002.
%\end{equation}
This discrepancy between predicted and measured value of $\alpha_s$ 
can be parametrized by $\epsilon_3 \equiv (\alpha_3(M_G) - \alpha_G )/ \alpha_G$.
%\begin{equation}
%\epsilon_3 \equiv \frac{ \alpha_3(M_G) - \alpha_G }{\alpha_G}.
%\end{equation}
The above mentioned results for $\alpha_s (M_Z)$ translate into $-5 \% < \epsilon_3 < -2.5 \% $. 
Let us see what we get in the region suggested by Yukawa unification.

The solid line 
in Fig.~\ref{figure:1} (left) represents $\epsilon_3$ as a function of $m_{16}$ with $\mu = M_{1/2} 
= m_A = 0.1 \ m_{16}$, $\tan \beta = 50$ 
and $A$ being linearly increased from $-1.7 \ m_{16}$ for $m_{16} = 1$ TeV to $-2.0 \
m_{16}$ for $m_{16} = 15$ TeV (to avoid negative stop mass squared which would occur if we took $A = 
-2m_{16}$ for $m_{16} \lesssim 2$ TeV with other parameters fixed as explained).
In the same figure (right) we see further variation of $\epsilon_3$ for 
one point from the plot on the left corresponding to $m_{16} = 5$ TeV. For different values 
of $\mu$ and $M_{1/2}$ (up to $0.2 \ m_{16}$), 
$\epsilon_3$ varies by $\sim 0.8 \%$ and this variation corresponds to dashed lines in 
the plot on the left. The shaded band correspond to $1 \sigma$ region for  gauge coupling 
unification. With our choice of parameters, the region $m_{16} \lesssim 2$ TeV is excluded 
due to chargino and/or stop masses being below experimental bounds.
For $m_{16} = 15$ TeV the mass of the stop is below 2 TeV and so it still might be  
acceptable. We  call the region of $m_{16} = [2, 15]$ TeV with other parameters 
varied as above the region with ``acceptable spectrum". Thus we see that in this region
$\epsilon_3$ varies from $-5 \%$ to $-1 \%$ and gets close to the band 
where gauge couplings unify within $1 \sigma$. 

\subsection{Scale of unification and proton decay}

It is well known, see for example Ref.~\cite{Dermisek:2000hr}, that the amplitude squared for
proton 
decay due to 
dimension 5 operators (mediated by color triplet higgsinos) scales 
approximately 
as 
$M_{1/2}^2/m_{16}^4$. Therefore, as far as SUSY 
spectrum is concerned, proton decay is maximally suppressed 
in the region required by Yukawa coupling unification.
%for small gaugino masses and 
%large squark and slepton masses. 

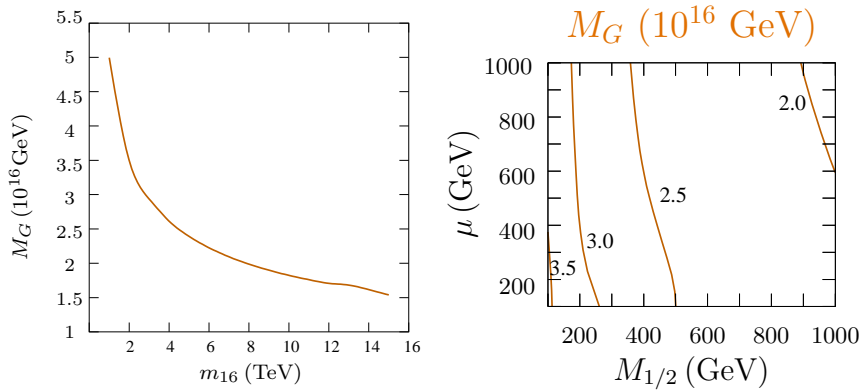
\begin{figure}[t]
\begin{center}
\input{M_G_p_70.pstex_t}
\caption{
Value of $M_G$ ($10^{16}$ GeV) as a function of $m_{16}$ with all other
parameters
fixed as explained in the text (left); and as a function of ($\mu$, $M_{1/2}$)
for $m_{16} = 5$ TeV and $A_0 = -1.8 \; m_{16}$ (right).}

\label{figure:2}
\end{center}
\end{figure}

Perhaps more interesting 
%and somewhat surprising 
thing related to proton decay is the value of 
the GUT scale itself. 
With increasing $m_{16}$ gauge couplings not only unify more precisely, they unify at 
lower scale. Fig.~\ref{figure:2} (left) shows $M_G$ as a function of $m_{16}$ with all other 
parameters fixed 
%in the same way as for the solid line 
as
in Fig.~\ref{figure:1}.  
In the region with ``acceptable spectrum" $M_G$ varies between $3.5 \times 10^{16}$ and $1.5 \times 
10^{16}$ GeV.  
In Fig.~\ref{figure:2} (right) we can see further variation of $M_G$  
for one point 
corresponding to $m_{16} = 5$ TeV. For different values of $\mu$ and $M_{1/2}$ 
it varies by $\sim 0.5 \times 10^{16}$ GeV so the curve in Fig.~\ref{figure:2} (left) 
can be shifted up and down by this amount. Comparing with Fig.~\ref{figure:1} we see that in 
the region where gauge couplings 
start to unify within $1 \sigma$, the 
corresponding value of the GUT scale is $M_G \simeq 1.0 \times 10^{16}$ GeV.
This however highly enhances proton decay due to dimension 6 operators (mediated by heavy 
gauge 
bosons). The amplitude squared is proportional to $1/(M_G)^4$. For $M_G = 1.0 \times 
10^{16}$ GeV the proton lifetime is $\sim 5 \times 10^{34}$ yr! This is just an order of magnitude 
from the current experimental bound and about 100 times larger than the lifetime obtained with 
usual assumption of $M_G \simeq 3.0 \times 10^{16}$ GeV corresponding to TeV scale SUSY 
breaking parameters.

\subsection{Conclusions}

I think it is highly non-trivial that there exists a region of SUSY parameter space which is 
simultaneously favored by gauge coupling unification, Yukawa coupling unification and proton 
decay. There is no reason why improving the situation with one feature of GUTs 
should not go in a wrong direction for another. Moreover, the same region is also favored 
when considering natural suppression of flavor and CP violation~\cite{Bagger:1999sy}, and can 
provide the 
right amount of neutralino relic density~\cite{Dermisek:2003vn}.
Perhaps the only thing which does not favor this region is the naturalness of EWSB.

Finally, it is interesting to note, that gauge coupling unification, Yukawa coupling unification  
and proton decay received some attention within higher dimensional GUTs. Sometimes problems associated 
with these features in four dimensional GUTs are used as a motivation for higher dimensional 
models~\cite{5dSO(10)}. 
Comparing to these, having somewhat heavy superpartners might not look so crazy anymore. :)

%\section*{Acknowledgments}
\subsection{Acknowledgments}
%\begin{acknowledgments}
I would like to thank K.~Tobe for comparing some numerical results and also 
T.~Bla\v zek, I.~Gogoladze, H. D. Kim, S.~Raby, K.~Tobe and J. D. Wells for useful comments and 
discussions on the subject.
This work was supported, in part, by the U.S.\ Department of Energy, Contract
DE-FG03-91ER-40674 and the Davis Institute for High Energy Physics.
%\end{acknowledgments}

%%%%%%%%%%%%%%%%%%%%%%%%%%%%%%%%%%%%%%%%%%%%%%%%%%%%%%%%%%%%%%%%%%%%%%%

\end{document}

%% file: eps_3_p_70.pstex_t
\begin{picture}(0,0)%
\includegraphics{eps_3_p_70.pstex}%
\end{picture}%
\setlength{\unitlength}{2763sp}%
\begingroup\makeatletter\ifx\SetFigFont\undefined%
\gdef\SetFigFont#1#2#3#4#5{%
  \reset@font\fontsize{#1}{#2pt}%
  \fontfamily{#3}\fontseries{#4}\fontshape{#5}%
  \selectfont}%
\fi\endgroup%
\begin{picture}(8324,3557)(1484,-4137)
\put(7722,-844){\makebox(0,0)[b]{\smash{{\SetFigFont{14}{16.8}{\rmdefault}{\mddefault}{\updefault}{\color[rgb]{1,0,0}$\epsilon_3 \, (\% )$}%
}}}}
\put(3099,-1027){\makebox(0,0)[b]{\smash{{\SetFigFont{7}{8.4}{\rmdefault}{\mddefault}{\updefault}{\color[rgb]{0,0,0}$\alpha_s = 0.1172 \mp 0.002$}%
}}}}
\put(6674,-3713){\makebox(0,0)[b]{\smash{{\SetFigFont{10}{12.0}{\familydefault}{\mddefault}{\updefault}{\color[rgb]{0,0,0}200}%
}}}}
\put(7273,-3713){\makebox(0,0)[b]{\smash{{\SetFigFont{10}{12.0}{\familydefault}{\mddefault}{\updefault}{\color[rgb]{0,0,0}400}%
}}}}
\put(7871,-3713){\makebox(0,0)[b]{\smash{{\SetFigFont{10}{12.0}{\familydefault}{\mddefault}{\updefault}{\color[rgb]{0,0,0}600}%
}}}}
\put(8469,-3713){\makebox(0,0)[b]{\smash{{\SetFigFont{10}{12.0}{\familydefault}{\mddefault}{\updefault}{\color[rgb]{0,0,0}800}%
}}}}
\put(9069,-3713){\makebox(0,0)[b]{\smash{{\SetFigFont{10}{12.0}{\familydefault}{\mddefault}{\updefault}{\color[rgb]{0,0,0}1000}%
}}}}
\put(6089,-3251){\makebox(0,0)[b]{\smash{{\SetFigFont{10}{12.0}{\familydefault}{\mddefault}{\updefault}{\color[rgb]{0,0,0}200}%
}}}}
\put(6089,-2735){\makebox(0,0)[b]{\smash{{\SetFigFont{10}{12.0}{\familydefault}{\mddefault}{\updefault}{\color[rgb]{0,0,0}400}%
}}}}
\put(6089,-2220){\makebox(0,0)[b]{\smash{{\SetFigFont{10}{12.0}{\familydefault}{\mddefault}{\updefault}{\color[rgb]{0,0,0}600}%
}}}}
\put(6089,-1704){\makebox(0,0)[b]{\smash{{\SetFigFont{10}{12.0}{\familydefault}{\mddefault}{\updefault}{\color[rgb]{0,0,0}800}%
}}}}
\put(6003,-1188){\makebox(0,0)[b]{\smash{{\SetFigFont{10}{12.0}{\familydefault}{\mddefault}{\updefault}{\color[rgb]{0,0,0}1000}%
}}}}
\put(7721,-4056){\makebox(0,0)[b]{\smash{{\SetFigFont{12}{14.4}{\familydefault}{\mddefault}{\updefault}{\color[rgb]{0,0,0}$M_{1/2} \, \rm (GeV)$}%
}}}}
\put(5659,-2220){\rotatebox{90.0}{\makebox(0,0)[b]{\smash{{\SetFigFont{12}{14.4}{\familydefault}{\mddefault}{\updefault}{\color[rgb]{0,0,0}$\mu \, \rm (GeV)$}%
}}}}}
\put(2160,-3797){\makebox(0,0)[b]{\smash{{\SetFigFont{7}{8.4}{\familydefault}{\mddefault}{\updefault}{\color[rgb]{0,0,0}0}%
}}}}
\put(2508,-3797){\makebox(0,0)[b]{\smash{{\SetFigFont{7}{8.4}{\familydefault}{\mddefault}{\updefault}{\color[rgb]{0,0,0}2}%
}}}}
\put(2859,-3797){\makebox(0,0)[b]{\smash{{\SetFigFont{7}{8.4}{\familydefault}{\mddefault}{\updefault}{\color[rgb]{0,0,0}4}%
}}}}
\put(3206,-3797){\makebox(0,0)[b]{\smash{{\SetFigFont{7}{8.4}{\familydefault}{\mddefault}{\updefault}{\color[rgb]{0,0,0}6}%
}}}}
\put(3558,-3797){\makebox(0,0)[b]{\smash{{\SetFigFont{7}{8.4}{\familydefault}{\mddefault}{\updefault}{\color[rgb]{0,0,0}8}%
}}}}
\put(3906,-3797){\makebox(0,0)[b]{\smash{{\SetFigFont{7}{8.4}{\familydefault}{\mddefault}{\updefault}{\color[rgb]{0,0,0}10}%
}}}}
\put(4256,-3797){\makebox(0,0)[b]{\smash{{\SetFigFont{7}{8.4}{\familydefault}{\mddefault}{\updefault}{\color[rgb]{0,0,0}12}%
}}}}
\put(4606,-3797){\makebox(0,0)[b]{\smash{{\SetFigFont{7}{8.4}{\familydefault}{\mddefault}{\updefault}{\color[rgb]{0,0,0}14}%
}}}}
\put(4954,-3797){\makebox(0,0)[b]{\smash{{\SetFigFont{7}{8.4}{\familydefault}{\mddefault}{\updefault}{\color[rgb]{0,0,0}16}%
}}}}
\put(3558,-4045){\makebox(0,0)[b]{\smash{{\SetFigFont{10}{12.0}{\familydefault}{\mddefault}{\updefault}{\color[rgb]{0,0,0}$m_{16} \,\rm (TeV)$}%
}}}}
\put(1966,-3608){\makebox(0,0)[b]{\smash{{\SetFigFont{7}{8.4}{\familydefault}{\mddefault}{\updefault}{\color[rgb]{0,0,0}-6}%
}}}}
\put(1966,-3228){\makebox(0,0)[b]{\smash{{\SetFigFont{7}{8.4}{\familydefault}{\mddefault}{\updefault}{\color[rgb]{0,0,0}-5}%
}}}}
\put(1966,-2789){\makebox(0,0)[b]{\smash{{\SetFigFont{7}{8.4}{\familydefault}{\mddefault}{\updefault}{\color[rgb]{0,0,0}-4}%
}}}}
\put(1966,-2349){\makebox(0,0)[b]{\smash{{\SetFigFont{7}{8.4}{\familydefault}{\mddefault}{\updefault}{\color[rgb]{0,0,0}-3}%
}}}}
\put(1966,-1973){\makebox(0,0)[b]{\smash{{\SetFigFont{7}{8.4}{\familydefault}{\mddefault}{\updefault}{\color[rgb]{0,0,0}-2}%
}}}}
\put(1966,-1153){\makebox(0,0)[b]{\smash{{\SetFigFont{7}{8.4}{\familydefault}{\mddefault}{\updefault}{\color[rgb]{0,0,0}0}%
}}}}
\put(1966,-712){\makebox(0,0)[b]{\smash{{\SetFigFont{7}{8.4}{\familydefault}{\mddefault}{\updefault}{\color[rgb]{0,0,0}1}%
}}}}
\put(1966,-1531){\makebox(0,0)[b]{\smash{{\SetFigFont{7}{8.4}{\familydefault}{\mddefault}{\updefault}{\color[rgb]{0,0,0}-1}%
}}}}
\put(1652,-2160){\rotatebox{90.0}{\makebox(0,0)[b]{\smash{{\SetFigFont{10}{12.0}{\familydefault}{\mddefault}{\updefault}{\color[rgb]{0,0,0}$\epsilon_3 \, (\%)$}%
}}}}}
\end{picture}%

%% file: M_G_p_70.pstex_t
\begin{picture}(0,0)%
\includegraphics{M_G_p_70.pstex}%
\end{picture}%
\setlength{\unitlength}{2763sp}%
\begingroup\makeatletter\ifx\SetFigFont\undefined%
\gdef\SetFigFont#1#2#3#4#5{%
  \reset@font\fontsize{#1}{#2pt}%
  \fontfamily{#3}\fontseries{#4}\fontshape{#5}%
  \selectfont}%
\fi\endgroup%
\begin{picture}(7842,3408)(1504,-3989)
\put(3669,-3901){\makebox(0,0)[b]{\smash{{\SetFigFont{8}{9.6}{\rmdefault}{\mddefault}{\updefault}{\color[rgb]{0,0,0}$m_{16} \, \rm (TeV)$}%
}}}}
\put(7623,-833){\makebox(0,0)[b]{\smash{{\SetFigFont{14}{16.8}{\rmdefault}{\mddefault}{\updefault}{\color[rgb]{.88,.44,0}$M_G$ ($10^{16}$ GeV)}%
}}}}
\put(1660,-2174){\rotatebox{90.0}{\makebox(0,0)[b]{\smash{{\SetFigFont{8}{9.6}{\familydefault}{\mddefault}{\updefault}{\color[rgb]{0,0,0}$M_G \, (10^{16} \rm GeV)$}%
}}}}}
\put(2043,-3238){\makebox(0,0)[b]{\smash{{\SetFigFont{7}{8.4}{\familydefault}{\mddefault}{\updefault}{\color[rgb]{0,0,0}1.5}%
}}}}
\put(2043,-3539){\makebox(0,0)[b]{\smash{{\SetFigFont{7}{8.4}{\familydefault}{\mddefault}{\updefault}{\color[rgb]{0,0,0}1}%
}}}}
\put(2043,-2937){\makebox(0,0)[b]{\smash{{\SetFigFont{7}{8.4}{\familydefault}{\mddefault}{\updefault}{\color[rgb]{0,0,0}2}%
}}}}
\put(2043,-2635){\makebox(0,0)[b]{\smash{{\SetFigFont{7}{8.4}{\familydefault}{\mddefault}{\updefault}{\color[rgb]{0,0,0}2.5}%
}}}}
\put(2043,-2333){\makebox(0,0)[b]{\smash{{\SetFigFont{7}{8.4}{\familydefault}{\mddefault}{\updefault}{\color[rgb]{0,0,0}3}%
}}}}
\put(2043,-2033){\makebox(0,0)[b]{\smash{{\SetFigFont{7}{8.4}{\familydefault}{\mddefault}{\updefault}{\color[rgb]{0,0,0}3.5}%
}}}}
\put(2043,-1731){\makebox(0,0)[b]{\smash{{\SetFigFont{7}{8.4}{\familydefault}{\mddefault}{\updefault}{\color[rgb]{0,0,0}4}%
}}}}
\put(2043,-1430){\makebox(0,0)[b]{\smash{{\SetFigFont{7}{8.4}{\familydefault}{\mddefault}{\updefault}{\color[rgb]{0,0,0}4.5}%
}}}}
\put(2043,-1069){\makebox(0,0)[b]{\smash{{\SetFigFont{7}{8.4}{\familydefault}{\mddefault}{\updefault}{\color[rgb]{0,0,0}5}%
}}}}
\put(2043,-707){\makebox(0,0)[b]{\smash{{\SetFigFont{7}{8.4}{\familydefault}{\mddefault}{\updefault}{\color[rgb]{0,0,0}5.5}%
}}}}
\put(6620,-3580){\makebox(0,0)[b]{\smash{{\SetFigFont{9}{10.8}{\familydefault}{\mddefault}{\updefault}{\color[rgb]{0,0,0}200}%
}}}}
\put(7194,-3580){\makebox(0,0)[b]{\smash{{\SetFigFont{9}{10.8}{\familydefault}{\mddefault}{\updefault}{\color[rgb]{0,0,0}400}%
}}}}
\put(7766,-3580){\makebox(0,0)[b]{\smash{{\SetFigFont{9}{10.8}{\familydefault}{\mddefault}{\updefault}{\color[rgb]{0,0,0}600}%
}}}}
\put(8339,-3580){\makebox(0,0)[b]{\smash{{\SetFigFont{9}{10.8}{\familydefault}{\mddefault}{\updefault}{\color[rgb]{0,0,0}800}%
}}}}
\put(8913,-3580){\makebox(0,0)[b]{\smash{{\SetFigFont{9}{10.8}{\familydefault}{\mddefault}{\updefault}{\color[rgb]{0,0,0}1000}%
}}}}
\put(6060,-3138){\makebox(0,0)[b]{\smash{{\SetFigFont{9}{10.8}{\familydefault}{\mddefault}{\updefault}{\color[rgb]{0,0,0}200}%
}}}}
\put(6060,-2644){\makebox(0,0)[b]{\smash{{\SetFigFont{9}{10.8}{\familydefault}{\mddefault}{\updefault}{\color[rgb]{0,0,0}400}%
}}}}
\put(6060,-2150){\makebox(0,0)[b]{\smash{{\SetFigFont{9}{10.8}{\familydefault}{\mddefault}{\updefault}{\color[rgb]{0,0,0}600}%
}}}}
\put(6060,-1656){\makebox(0,0)[b]{\smash{{\SetFigFont{9}{10.8}{\familydefault}{\mddefault}{\updefault}{\color[rgb]{0,0,0}800}%
}}}}
\put(5977,-1162){\makebox(0,0)[b]{\smash{{\SetFigFont{9}{10.8}{\familydefault}{\mddefault}{\updefault}{\color[rgb]{0,0,0}1000}%
}}}}
\put(7622,-3908){\makebox(0,0)[b]{\smash{{\SetFigFont{12}{14.4}{\familydefault}{\mddefault}{\updefault}{\color[rgb]{0,0,0}$M_{1/2} \, \rm (GeV)$}%
}}}}
\put(5648,-2150){\rotatebox{90.0}{\makebox(0,0)[b]{\smash{{\SetFigFont{12}{14.4}{\familydefault}{\mddefault}{\updefault}{\color[rgb]{0,0,0}$\mu \, \rm (GeV)$}%
}}}}}
\end{picture}%